\documentclass[12pt]{article}
\usepackage[english]{babel}
\usepackage{bm}
\usepackage{fullpage}
\usepackage{cite}

\usepackage{newpxtext,newpxmath}

\usepackage[utf8]{inputenc}
\usepackage{amsthm}
\usepackage{amssymb}
\usepackage{amsmath}
\usepackage{bbold}
\usepackage{bbm}
\usepackage[pdftex, backref=page]{hyperref}
\usepackage{braket}
\usepackage{dsfont}
\usepackage{mathdots}
\usepackage{mathtools}
\usepackage{enumerate}
\usepackage[shortlabels]{enumitem}
\usepackage{csquotes}
\usepackage{stmaryrd}
\usepackage[cal=boondox]{mathalfa}
\usepackage{graphicx}
\usepackage{stackengine}
\usepackage{scalerel}
\usepackage{tensor}       
\usepackage{array}
\usepackage{makecell}
\newcolumntype{x}[1]{>{\centering\arraybackslash}p{#1}}
\usepackage{tikz}
\usepackage{pgfplots}
\usetikzlibrary{shapes.geometric, shapes.misc, positioning, arrows, arrows.meta, decorations.pathreplacing, decorations.pathmorphing, patterns, angles, quotes, calc}
\usepackage{booktabs}
\usepackage{xfrac}
\usepackage{siunitx}
\usepackage{centernot}
\usepackage{comment}
\usepackage{chngcntr}
\usepackage{caption}
\usepackage{subcaption}

\newtheorem{thm}{Theorem}
\newtheorem*{thm*}{Theorem}

\newtheorem*{prop*}{Proposition}
\newtheorem{lemma}[thm]{Lemma}
\newtheorem*{lemma*}{Lemma}

\newtheorem*{cor*}{Corollary}

\newtheorem*{cj*}{Conjecture}
\newtheorem{Def}[thm]{Definition}
\newtheorem*{Def*}{Definition}

\newtheorem*{question*}{Question}

\newtheorem*{problem*}{Problem}
\newtheorem*{example*}{Example}

\makeatletter
\def\thmhead@plain#1#2#3{%
  \thmname{#1}\thmnumber{\@ifnotempty{#1}{ }\@upn{#2}}%
  \thmnote{ {\the\thm@notefont#3}}}
\let\thmhead\thmhead@plain
\makeatother

\theoremstyle{definition}
\newtheorem{rem}[thm]{Remark}

\newcommand{\bb}{\begin{equation}\begin{aligned}\hspace{0pt}}
\newcommand{\bbb}{\begin{equation*}\begin{aligned}}
\newcommand{\ee}{\end{aligned}\end{equation}}
\newcommand{\eee}{\end{aligned}\end{equation*}}

\renewcommand{\epsilon}{\varepsilon}

\DeclareMathAlphabet{\pazocal}{OMS}{zplm}{m}{n}

\newcommand{\lsmatrix}{\left(\begin{smallmatrix}}
\newcommand{\rsmatrix}{\end{smallmatrix}\right)}

\stackMath

\stackMath

\makeatletter
\newcommand*\rel@kern[1]{\kern#1\dimexpr\macc@kerna}
\newcommand*\widebar[1]{%
  \begingroup
  \def\mathaccent##1##2{%
    \rel@kern{0.8}%
    \overline{\rel@kern{-0.8}\macc@nucleus\rel@kern{0.2}}%
    \rel@kern{-0.2}%
  }%
  \macc@depth\@ne
  \let\math@bgroup\@empty \let\math@egroup\macc@set@skewchar
  \mathsurround\z@ \frozen@everymath{\mathgroup\macc@group\relax}%
  \macc@set@skewchar\relax
  \let\mathaccentV\macc@nested@a
  \macc@nested@a\relax111{#1}%
  \endgroup
}

\counterwithin*{equation}{part}
\counterwithin*{thm}{part}
\counterwithin*{figure}{part}

\tikzset{meter/.append style={draw, inner sep=10, rectangle, font=\vphantom{A}, minimum width=30, line width=.8, path picture={\draw[black] ([shift={(.1,.3)}]path picture bounding box.south west) to[bend left=50] ([shift={(-.1,.3)}]path picture bounding box.south east);\draw[black,-latex] ([shift={(0,.1)}]path picture bounding box.south) -- ([shift={(.3,-.1)}]path picture bounding box.north);}}}
\tikzset{roundnode/.append style={circle, draw=black, fill=gray!20, thick, minimum size=10mm}}
\tikzset{squarenode/.style={rectangle, draw=black, fill=none, thick, minimum size=10mm}}

\definecolor{Blues5seq1}{RGB}{239,243,255}
\definecolor{Blues5seq2}{RGB}{189,215,231}
\definecolor{Blues5seq3}{RGB}{107,174,214}
\definecolor{Blues5seq4}{RGB}{49,130,189}
\definecolor{Blues5seq5}{RGB}{8,81,156}

\definecolor{Greens5seq1}{RGB}{237,248,233}
\definecolor{Greens5seq2}{RGB}{186,228,179}
\definecolor{Greens5seq3}{RGB}{116,196,118}
\definecolor{Greens5seq4}{RGB}{49,163,84}
\definecolor{Greens5seq5}{RGB}{0,109,44}

\definecolor{Reds5seq1}{RGB}{254,229,217}
\definecolor{Reds5seq2}{RGB}{252,174,145}
\definecolor{Reds5seq3}{RGB}{251,106,74}
\definecolor{Reds5seq4}{RGB}{222,45,38}
\definecolor{Reds5seq5}{RGB}{165,15,21}

\allowdisplaybreaks

\usepackage[most,breakable]{tcolorbox}
	{\expandafter\ifstrequal\expandafter{#1}{orange}{\begin{tcolorbox}[colback=red!15,colframe=orange!15,breakable,enhanced]}{\begin{tcolorbox}[colback=Blues5seq1,colframe=Blues5seq5,breakable,enhanced]}}%
	{\end{tcolorbox}}

\usepackage{authblk}

\newtheorem*{remark}{Remark}

\allowdisplaybreaks[1] 
\usepackage{graphicx,import,xcolor,transparent}
\usepackage{centernot}

\begin{document}

\title{Universal state merging for almost i.i.d.\ states}

\author{Nilanjana Datta}

\affil{
Department of Applied Mathematics and Theoretical Physics\\
University of Cambridge\\
Cambridge CB3 0WA, United Kingdom
}

\date{}

\maketitle

\begin{abstract}
{
In quantum state merging, Alice aims to transfer her share of a
bipartite state to Bob while preserving the joint state with a
purifying reference. For i.i.d.\ copies of $\rho_{AB}$, the optimal
asymptotic net entanglement cost is $S(A|B)_\rho$ ebits per pair.
We show that this cost is robust under deviations from exact i.i.d.\
structure. For every $R>S(A|B)_\rho$, a single sequence of one-way
LOCC protocols, depending only on $\rho_{AB}$ and $R$, merges every
weakly almost i.i.d.\ source along $\rho_{AB}$ with vanishing error.
The net entanglement cost is at most $nR$ ebits for all sufficiently
large blocklengths $n$. This guarantee also covers the Wasserstein
and Mazzola--Sutter--Renner subclasses. Each class contains the
exact i.i.d.\ source, so $S(A|B)_\rho$ is optimal for universal
merging within each class. More generally, the same protocols
achieve vanishing error whenever the two-pair marginal of the
permutation-averaged source converges in trace norm to
$\rho_{AB}^{\otimes 2}$.}
\end{abstract}

\section{Introduction}

How much entanglement does Alice need to transfer her share of a
quantum state to Bob? In state merging, Alice and Bob share a
bipartite state $\rho_{AB}$, and the aim is to transfer Alice's
register $A$ to Bob while preserving the joint state with a
purifying reference. Bob uses his register $B$ as quantum side
information. {Alice and Bob use local operations and
classical communication (LOCC), with messages sent from Alice to
Bob, and may draw on shared entanglement.} The net entanglement
cost is the number of ebits shared at the start minus the number
remaining at the end.

{State merging was originally studied in the asymptotic
i.i.d.\ setting. Alice and Bob share $\rho_{AB}^{\otimes n}$ and
seek to minimise the net entanglement cost per copy in the limit
$n\to\infty$, with vanishing merging error.}
For notational simplicity, we write iid for i.i.d.\ throughout
the rest of the paper. Horodecki, Oppenheim and Winter~\cite{HOW}
showed that the optimal cost is $S(A|B)_\rho$ ebits per copy, where $S(A|B)_\rho$ is the conditional entropy of $A$ given $B$
in the state $\rho_{AB}$.
When $S(A|B)_\rho<0$, merging yields a net gain of entanglement.

This cost can also be achieved using the fully
quantum Slepian--Wolf protocol of Abeyesinghe, Devetak, Hayden
and Winter~\cite{ADHW}. For a purification $\lvert\psi\rangle_{ABZ}$ of $\rho_{AB}$,
the asymptotic rates of quantum communication and entanglement
generation are $\frac12 I(A:Z)_\psi$ qubits and
$\frac12 I(A:B)_\psi$ ebits per input copy,
respectively.\footnote{For a bipartite state $\tau_{XY}$,
the quantum mutual information is
$I(X:Y)_\tau=S(\tau_X)+S(\tau_Y)-S(\tau_{XY})$.}
Teleporting the transmitted qubits gives a one-way LOCC
protocol with net entanglement cost
$\frac12 I(A:Z)_\psi-\frac12 I(A:B)_\psi=S(A|B)_\rho$
per copy.

In the one-shot setting, Berta~\cite{Berta2009} bounded the
entanglement cost of state merging in terms of smooth conditional
entropies. Dupuis et al.~\cite{DupuisBertaWullschlegerRenner2014}
subsequently derived achievability and converse bounds using
one-shot decoupling. For compound iid sources,
Bjelakovi\'c, Boche and Jan{\ss}en~\cite{BjelakovicBocheJanssen2013}
established universal merging, where a single sequence of
protocols works for every tensor-power source generated by a
state in a known set.

The iid assumption specifies the global state shared by Alice
and Bob to be an exact tensor power. {Even a complete
tomography of all marginals of a fixed size cannot generally 
certify this global structure. This leads to the following natural question: {\em{Does the iid entanglement cost
remain achievable under a local iid condition that allows
long-range correlations?}}}

We address this question for \emph{almost iid states}. Several
ways to formalise departures from exact iid structure for quantum
sources and channels have been studied in the recent
literature~\cite{mazzolasutterrenner1,
girardi2026newapproachesiidinformation,MazzolaSutterRenner2026Stein,
GirardiDattaDePalmaLami,datta2026entropy,
Datta2026EntanglementManipulation,GirardiLeeHayashiLami2026,
RizzoEtAl2026,YeBerghDatta2026DeFinetti,YeBerghDatta2026Channels}. Three notions are relevant here. Mazzola, Sutter and Renner
introduced the MSR class, and Girardi, De Palma and Lami
introduced the Wasserstein and weakly almost iid classes.
The latter authors showed that, along a fixed reference state,
the MSR class is contained in the Wasserstein class, which in
turn is contained in the weakly almost iid class. We establish
robustness of state merging for the broadest class, so our
result applies to all three.

For a weakly almost iid source along $\rho_{AB}$ and each fixed
$k$, the trace distance between a $k$-pair marginal and
$\rho_{AB}^{\otimes k}$ tends to zero as $n\to\infty$, on average
over a uniform choice of the $k$ pairs.
See Definition~\ref{def:waiid} in Section~\ref{sec:waiid}.
This asymptotic local iid behaviour is compatible with
arbitrarily long-range classical or quantum correlations,
including multipartite
entanglement~\cite{girardi2026newapproachesiidinformation,datta2026entropy}.

{
\medskip
\noindent
\textbf{Main result.}
{The optimal asymptotic net entanglement cost for
universal state merging of weakly almost iid sources along
$\rho_{AB}$ is $S(A|B)_\rho$ ebits per pair
(Theorem~\ref{thm:universal-waiid-merging}).}
For every $R>S(A|B)_\rho$, a single sequence
of one-way LOCC protocols, depending only on $\rho_{AB}$ and
$R$, merges every source in this class with vanishing error
and net entanglement cost at most $nR$ for all sufficiently
large $n$. {The same protocols also work whenever
the marginal on the first two pairs of the permutation-averaged
block state converges in trace norm to $\rho_{AB}^{\otimes 2}$.
The iid converse gives the matching lower bound, since the
class contains the exact iid source along $\rho_{AB}$.}

\medskip
\noindent
\textbf{Proof sketch.}

The proof uses two ingredients. The first is the domination
lemma (Lemma~\ref{lem:product-state-domination}).
{To prove it, we intersect the range of the universal
typical projector from Lemma~3 of~\cite{datta2026entropy}
with the direct sum of selected Schur sectors.} The Schur-spectrum concentration estimate of Rizzo et
al.~\cite{RizzoEtAl2026} shows that the selected sectors carry
asymptotically all the weight of the permutation-averaged source.
On those sectors, permutation invariance and their bound on
the dimensions of irreducible permutation representations give
an upper bound on the eigenvalues of the source. The spectral
cutoff also retains asymptotically all the source weight and
gives a lower bound on the eigenvalues of a full-rank tensor-power
state on its range. On the intersection, these bounds give an
operator inequality between the unnormalised restriction of the
permutation-averaged source and a scalar multiple of the
tensor-power state.}

{The second ingredient is a uniform exponential
error bound for iid merging
(Lemma~\ref{lem:exponential-iid-merging}), obtained from the
compound-source result of Boche and Jan{\ss}en~\cite{BocheJanssen2014}.
Using isometric descriptions of the actual and ideal merging maps,
we bound the squared merging error by the expectation of a fixed
positive operator $X_n$ on the permutation-averaged source tensored
with the initial entanglement resource. The iid error bound makes
the corresponding expectation exponentially small on the full-rank
tensor-power input. By positivity of $X_n$, the domination inequality
bounds the expectation on the retained part of the source by the
iid expectation multiplied by the scalar factor. Uniformity lets us
fix the error exponent first. We then choose the full-rank single-pair
state close enough to $\rho_{AB}$ and the tolerance in the domination
bound small enough that the scalar factor's growth exponent is below
the error decay exponent. The contribution from the complement
vanishes as its weight tends to zero.}

\section{Weakly almost iid states}\label{sec:waiid}

All Hilbert spaces are finite-dimensional. We write $\mathcal H_A$
for the Hilbert space of a system $A$ and
$\mathcal H_{AB}=\mathcal H_A\otimes\mathcal H_B$ for the joint
Hilbert space of $A$ and $B$. We denote the set of density
operators on $\mathcal H$ by $\mathcal D(\mathcal H)$. A density
operator is a positive semidefinite operator of unit trace.
A pure state is a rank-one projection
$\lvert\psi\rangle\langle\psi\rvert$, where $\lvert\psi\rangle$
is a unit vector. We sometimes write $\psi$ for this projection.

For an operator $X$ on $\mathcal H$, its trace norm is
$\|X\|_1=\operatorname{Tr}\sqrt{X^\dagger X}$.
For $\rho,\sigma\in\mathcal D(\mathcal H)$, the trace distance
and fidelity are defined by
\bb
T(\rho,\sigma)=\frac12\|\rho-\sigma\|_1,
\qquad
F(\rho,\sigma)=\bigl\|\sqrt{\rho}\sqrt{\sigma}\bigr\|_1.
\ee
With this convention, for a pure state
$\psi=\lvert\psi\rangle\langle\psi\rvert$ we have
$F(\rho,\psi)=\sqrt{\langle\psi\rvert\rho\lvert\psi\rangle}$.

All logarithms are to base~$2$. The von Neumann entropy of a
state $\rho$ is $S(\rho)=-\operatorname{Tr}(\rho\log\rho)$,
and the Shannon entropy of a probability vector $p$ is
$H(p)=-\sum_i p_i\log p_i$. We use the convention $0\log 0=0$.
For a bipartite state $\rho_{AB}$, the conditional entropy is
$S(A|B)_\rho=S(\rho_{AB})-S(\rho_B)$, where
$\rho_B=\operatorname{Tr}_A\rho_{AB}$.

For states $\rho$ and $\sigma$ on the same Hilbert space, the
quantum relative entropy is
$D(\rho\Vert\sigma)=\operatorname{Tr}[\rho(\log\rho-\log\sigma)]$
if $\operatorname{supp}(\rho)\subseteq\operatorname{supp}(\sigma)$,
and $D(\rho\Vert\sigma)=+\infty$ otherwise. The support of a state
is the span of its eigenvectors with positive eigenvalues.
In the trace formula, each logarithm is evaluated on the support
of its argument and extended by zero on its orthogonal complement.

We write $\|X\|_\infty$ for the operator norm. For self-adjoint
operators $X$ and $Y$, we write $X\le Y$ if $Y-X$ is positive
semidefinite.

{We use the definition of a weakly almost iid source
due to Girardi, De Palma and
Lami~\cite[Definition~11]{girardi2026newapproachesiidinformation},
with each site taken to be a bipartite pair $AB$.}

\begin{Def}[({Weakly almost iid bipartite source})]\label{def:waiid}
{
Let $\rho=\rho_{AB}\in\mathcal D(\mathcal H_{AB})$ be a fixed
reference state. A sequence $\bm{\omega}=(\omega_n)_{n\ge 1}$,
with $\omega_n\in\mathcal D(\mathcal H_{AB}^{\otimes n})$, is called
a weakly almost iid bipartite quantum source along $\rho$ if,
for every fixed positive integer $t$,
\bb\label{eq:walmost iid}
\lim_{n\to\infty}
\mathbb E_{\substack{I\subseteq[n]\\|I|=t}}
\bigl\|(\omega_n)_I-\rho^{\otimes t}\bigr\|_1
=0.
\ee
Here $[n]=\{1,\ldots,n\}$. For $n\ge t$, the expectation is
uniform over all $t$-element subsets of $[n]$. The state
$(\omega_n)_I$ is the marginal of $\omega_n$ on the pairs
$A_iB_i$ with $i\in I$, ordered by increasing index.}
\end{Def}

This local marginal condition is compatible with long-range
classical correlations and multipartite entanglement, as
illustrated by the examples in Section~3.3
of~\cite{girardi2026newapproachesiidinformation} and Examples~1--3
of~\cite{datta2026entropy}.

{

\section{Product-state domination from two-site marginals}

\begin{lemma}[(Product-state domination on a high-probability subspace)]
\label{lem:product-state-domination}
Let $\rho$ be a state on a $d$-dimensional Hilbert space
$\mathcal H$. Fix $q\in(0,1)$ and $\eta>0$, and set
$\rho_q=(1-q)\rho+qI/d$. There exists a sequence of orthogonal
projectors $(Q_n)_{n\ge 1}$ on $\mathcal H^{\otimes n}$,
depending only on $\rho$, $q$ and $\eta$, such that each $Q_n$
commutes with all tensor-factor permutations.

For a sequence of states $(\omega_n)_{n\ge 1}$ on
$\mathcal H^{\otimes n}$, let
\bb
\bar\omega_n=\frac{1}{n!}\sum_{\pi\in S_n}
V_\pi\omega_n V_\pi^\dagger,
\ee
where $V_\pi$ permutes the tensor factors according to $\pi$.
{For $n\ge 2$, write $(\bar\omega_n)_{12}$ for the
marginal on the first two sites.} If
\bb
\bigl\|(\bar\omega_n)_{12}-\rho^{\otimes 2}\bigr\|_1
\longrightarrow 0 \quad {\hbox{as}} \quad n \to \infty,
\ee
then
\bb
\operatorname{Tr}(Q_n\omega_n)
=\operatorname{Tr}(Q_n\bar\omega_n)\longrightarrow 1.
\ee
{For each $n\ge 1$ and every state $\omega_n$ on
$\mathcal H^{\otimes n}$, the same $Q_n$ satisfies}
\bb
Q_n\bar\omega_nQ_n
\le (n+d)^{c_d}
2^{n[D(\rho\Vert\rho_q)+\eta]}\rho_q^{\otimes n},
\qquad
c_d=\frac{(d-1)(d+2)}{2}.
\ee
\end{lemma}

\begin{proof}
{We compare the operators on the intersection of
the ranges of two commuting projectors. On the range of the
Schur projector $F_n$, we bound $\bar\omega_n$ from above by a
multiple of the identity. On the range of the spectral projector
$P_n$, we bound $\rho_q^{\otimes n}$ from below by a multiple
of the identity. The two-site condition ensures that both
ranges carry asymptotically all the weight of $\bar\omega_n$.}

Set $\varepsilon=\eta/2$.
Let $p=(p_1,\ldots,p_d)$
be the eigenvalue vector of $\rho$ in non-increasing order.
For probability vectors $a$ and $p$, write
$d_{\mathrm{TV}}(a,p)=\frac12\sum_i |a_i-p_i|$.
Since $H(p)=S(\rho)$, continuity of the Shannon entropy gives
a $\delta>0$, depending only on $p$ and $\varepsilon$, such that
\bb
H(a)\ge S(\rho)-\varepsilon
\qquad\text{whenever}\qquad
d_{\mathrm{TV}}(a,p)\le\delta.
\ee

\medskip
\noindent\textbf{The Schur projector:}\quad
Let $X$ denote the single-site system, with
$\mathcal H_X=\mathcal H$. Write the Schur--Weyl decomposition as
\bb
\mathcal H_X^{\otimes n}
\simeq
\bigoplus_{\lambda\in Y_{n,d}}
\mathcal U_{X,\lambda}\otimes\mathcal V_{X,\lambda},
\ee
where $Y_{n,d}$ is the set of Young diagrams of size $n$
with at most $d$ rows. The permutation representation acts
trivially on $\mathcal U_{X,\lambda}$ and irreducibly on
$\mathcal V_{X,\lambda}$, with inequivalent irreducible
representations for distinct $\lambda$. Let $\Pi_\lambda$
be the projector onto the summand indexed by $\lambda$.
We identify $\lambda$ with its vector of row lengths,
padded with zeros to length $d$.

For an iid input $\rho^{\otimes n}$, the normalised Schur
label {converges in probability to the ordered
spectrum $p$ of $\rho$~\cite{KeylWerner}. We retain the
Schur blocks whose normalised labels lie within total
variation distance $\delta$ of $p$, setting}
\bb
F_n=
\sum_{\substack{\lambda\in Y_{n,d}\\
                d_{\mathrm{TV}}(\lambda/n,p)\le\delta}}
\Pi_\lambda.
\ee
Let $\Lambda_n$ denote the random outcome of measuring
$\bar\omega_n$ with the projectors $\{\Pi_\lambda\}_\lambda$,
so that
$\Pr(\Lambda_n=\lambda)=\operatorname{Tr}(\Pi_\lambda\bar\omega_n)$.
{For $n\ge 2$, the average of the ordered two-site
marginals of $\bar\omega_n$ equals $(\bar\omega_n)_{12}$ by
permutation invariance. Applying Lemma~27 of
Rizzo et al.~\cite{RizzoEtAl2026} to $\bar\omega_n$ gives}
\bb
\mathbb E\left\|\frac{\Lambda_n}{n}-p\right\|_2^2
\le
\frac{d-\operatorname{Tr}(\rho^2)}{n}
+
3\bigl\|(\bar\omega_n)_{12}-\rho^{\otimes 2}\bigr\|_1.
\ee
Here $\|\cdot\|_2$ is the Euclidean norm. By Cauchy--Schwarz,
$d_{\mathrm{TV}}(a,p)^2\le (d/4)\|a-p\|_2^2$. Markov's
inequality and the two-site assumption therefore give
\bb
\operatorname{Tr}\bigl((I-F_n)\bar\omega_n\bigr)
=
\Pr\!\left\{d_{\mathrm{TV}}(\Lambda_n/n,p)>\delta\right\}
\le
\frac{d}{4\delta^2}
\mathbb E\left\|\frac{\Lambda_n}{n}-p\right\|_2^2
\longrightarrow 0.
\ee

Since $\bar\omega_n$ commutes with all permutation unitaries,
Schur's lemma gives
\bb
\bar\omega_n
=
\bigoplus_{\lambda\in Y_{n,d}}
Z_\lambda\otimes
\frac{I_{\mathcal V_{X,\lambda}}}{\dim\mathcal V_{X,\lambda}},
\ee
where $Z_\lambda\ge 0$ acts on $\mathcal U_{X,\lambda}$ and
$\sum_{\lambda}\operatorname{Tr} Z_\lambda=1$.
{The largest eigenvalue of $Z_\lambda$ is at most
its trace, which is at most one. The maximally mixed factor
on $\mathcal V_{X,\lambda}$ divides each eigenvalue of
$Z_\lambda$ by $\dim\mathcal V_{X,\lambda}$. Thus every
eigenvalue of the $\lambda$ block is at most
$1/\dim\mathcal V_{X,\lambda}$.} Lemma~17 of Rizzo et al.~\cite{RizzoEtAl2026} gives
\bb
\dim\mathcal V_{X,\lambda}
\ge (n+d)^{-c_d}2^{nH(\lambda/n)}.
\ee
{
For each block retained by $F_n$, we have
$d_{\mathrm{TV}}(\lambda/n,p)\le\delta$, so the entropy
continuity bound above gives $H(\lambda/n)\ge S(\rho)-\varepsilon$.
Taking the reciprocal of the dimension bound on these blocks
gives
\bb
\label{eq:domination-schur-bound}
F_n\bar\omega_nF_n
\le (n+d)^{c_d}2^{-n[S(\rho)-\varepsilon]}F_n.
\ee}
\medskip
\noindent\textbf{The spectral cutoff:}\quad
Since $\rho_q\ge(q/d)I$, the observable $A_q=-\log\rho_q$
is bounded. Set $h_q=\operatorname{Tr}(\rho A_q)$ and define
\bb
\bar A_n=\frac1n\sum_{i=1}^n A_q^{(i)}
=-\frac1n\log(\rho_q^{\otimes n}),
\qquad
P_n=\{\bar A_n\le (h_q+\varepsilon)I\},
\ee
where braces denote the spectral projector onto the
eigenspaces satisfying the indicated inequality.
{
Since $\rho_q^{\otimes n}=2^{-n\bar A_n}$ and
$x\mapsto 2^{-nx}$ is decreasing, we can write the same cutoff as
\bb
P_n=\{\rho_q^{\otimes n}\ge 2^{-n(h_q+\varepsilon)}I\}.
\ee
}

{The projector $P_n$ is the spectral cutoff used in Lemma~3
of~\cite{datta2026entropy}. To establish its high-weight property
under the two-site condition, we bound the second moment of
$\bar A_n$ about the reference mean $h_q$. 
Set $B=A_q-h_qI$. For $n\ge 2$, expanding the square and using
permutation invariance of $\bar\omega_n$ gives
}
\bb
\begin{aligned}
\operatorname{Tr}\!\left[\bar\omega_n(\bar A_n-h_qI)^2\right]
&=
\frac1n\operatorname{Tr}\!\left[(\bar\omega_n)_1 B^2\right]\\
&\quad+
\frac{n-1}{n}
\operatorname{Tr}\!\left[(\bar\omega_n)_{12}(B\otimes B)\right].
\end{aligned}
\ee
The first term is at most $\|B\|_\infty^2/n$. By the two-site
assumption, the second term tends to
$\operatorname{Tr}[\rho^{\otimes 2}(B\otimes B)]
=(\operatorname{Tr}(\rho B))^2=0$.
{On the range of $I-P_n$, the eigenvalues of
$\bar A_n-h_qI$ exceed $\varepsilon$, so}
\bb
I-P_n\le \varepsilon^{-2}(\bar A_n-h_qI)^2.
\ee
{Taking the expectation under $\bar\omega_n$ now
gives} $\operatorname{Tr}((I-P_n)\bar\omega_n)\to 0$.

\medskip
\noindent\textbf{Intersection and domination:}\quad
The operator $\rho_q^{\otimes n}$ commutes with all permutation
unitaries, so it is block diagonal in the Schur--Weyl
decomposition. Thus $P_n$ commutes with each $\Pi_\lambda$
and hence with $F_n$. Define $Q_n=F_nP_n$, the orthogonal
projector onto the intersection of their ranges.
\bb
I-Q_n\le (I-F_n)+(I-P_n).
\ee
{The weight bounds for $F_n$ and $P_n$ therefore give}
$\operatorname{Tr}(Q_n\bar\omega_n)\to 1$. Both $F_n$ and
$P_n$ commute with all tensor-factor permutations, so $Q_n$
does as well. Consequently,
$\operatorname{Tr}(Q_n\omega_n)=\operatorname{Tr}(Q_n\bar\omega_n)$.
The construction depends only on $\rho$, $q$ and $\eta$.

Since $Q_n\le P_n$, the cutoff gives
\bb
\rho_q^{\otimes n}\ge 2^{-n(h_q+\varepsilon)}P_n
\ge 2^{-n(h_q+\varepsilon)}Q_n.
\ee
{Compressing \eqref{eq:domination-schur-bound} by
$Q_n$ gives the first inequality below, since $Q_nF_n=F_nQ_n=Q_n$.
The second follows from the spectral lower bound.}
\bb
\begin{aligned}
Q_n\bar\omega_nQ_n
&\le (n+d)^{c_d}2^{-n[S(\rho)-\varepsilon]}Q_n\\
&\le (n+d)^{c_d}
2^{n[h_q-S(\rho)+2\varepsilon]}\rho_q^{\otimes n}.
\end{aligned}
\ee
The claimed bound follows from
$h_q-S(\rho)=D(\rho\Vert\rho_q)$ and $2\varepsilon=\eta$.
\end{proof}

\begin{remark}
Every weakly almost iid source along $\rho$ satisfies the
two-site condition in Lemma~\ref{lem:product-state-domination}.
{For $n\ge 2$, each ordered pair of distinct sites
occurs in the first two positions of exactly $(n-2)!$
permutations. Writing $(\omega_n)_{ij}$ for the marginal
with its factors ordered as $i,j$, we therefore have}
\bb
(\bar\omega_n)_{12}
=\frac{1}{n(n-1)}\sum_{i\ne j}(\omega_n)_{ij}.
\ee
Since $\rho^{\otimes 2}$ is invariant under swapping its two
factors, convexity and unitary invariance of the trace norm give
\bb
\bigl\|(\bar\omega_n)_{12}-\rho^{\otimes 2}\bigr\|_1
\le
\frac{1}{\binom n2}\sum_{i<j}
\bigl\|(\omega_n)_{ij}-\rho^{\otimes 2}\bigr\|_1
\longrightarrow 0,
\ee
where the convergence follows from the weakly almost iid
assumption.
\end{remark}

\section{One-way state merging}
\label{subsec:state-merging}

Suppose Alice and Bob share a state $\omega_{A^nB^n}$, with
Alice holding $A^n$ and Bob holding $B^n$. Let
$\psi_{A^nB^nZ}$ be a purification of $\omega_{A^nB^n}$,
where $Z$ is an inaccessible reference system. The aim of
state merging is to transfer Alice's share to Bob while
preserving the joint state with the reference.

For each positive integer $M$, let $\Phi_M$ denote the density
operator of the maximally entangled vector
\bb
\lvert\Phi_M\rangle
=
\frac{1}{\sqrt{M}}\sum_{j=1}^M
\lvert j\rangle\otimes\lvert j\rangle.
\ee
Alice and Bob initially share $\Phi_{K_n}$ on registers
$A_0$ and $B_0$, respectively. They apply a one-way LOCC
protocol, with classical communication from Alice to Bob.
Bob's output registers for the source are $\widehat A^n$ and
$B^n$, where $\widehat A^n$ is isomorphic to $A^n$. The
prescribed final entangled state is $\Phi_{L_n}$ on $A_1B_1$,
with $A_1$ held by Alice and $B_1$ by Bob. The ranks $K_n$
and $L_n$ are fixed by the protocol. Its net entanglement cost
is $\log K_n-\log L_n$ ebits. A negative cost corresponds
to a net gain of shared entanglement.

Let $\Omega_n$ be the joint output on
$\widehat A^n B^n A_1B_1Z$ when the protocol acts on
$A^nB^nA_0B_0$ in the state
$\psi_{A^nB^nZ}\otimes\Phi_{K_n}^{A_0B_0}$, with the
identity channel acting on $Z$.

To describe the ideal output, fix identifications of $A^n$
with $\widehat A^n$ and of $A_0B_0$ with an environment
register $G_n$. Define the ideal merging isometry
\bb
W_n:\mathcal H_{A^nB^nA_0B_0}
\longrightarrow
\mathcal H_{\widehat A^n B^n A_1B_1G_n}
\ee
on product vectors by
\bb
W_n\bigl(\lvert v\rangle_{A^nB^n}
          \otimes\lvert u\rangle_{A_0B_0}\bigr)
=
\lvert v\rangle_{\widehat A^n B^n}
\otimes\lvert\Phi_{L_n}\rangle_{A_1B_1}
\otimes\lvert u\rangle_{G_n},
\ee
and extend it linearly. The relabelling of the input registers
and the unit norm of $\lvert\Phi_{L_n}\rangle$ ensure that
$W_n$ is an isometry. Tracing out $G_n$ gives the ideal output
\bb
\begin{aligned}
\Omega_n^{\mathrm{ideal}}
&=
\operatorname{Tr}_{G_n}\!\left[
(W_n\otimes I_Z)
\bigl(\psi_{A^nB^nZ}\otimes\Phi_{K_n}^{A_0B_0}\bigr)
(W_n^\dagger\otimes I_Z)
\right]\\
&=
\psi_{\widehat A^n B^n Z}\otimes\Phi_{L_n}^{A_1B_1},
\end{aligned}
\ee
where $\psi_{\widehat A^n B^n Z}$ is the input purification
with $A^n$ relabelled as $\widehat A^n$.

The state merging error is
\bb
T(\Omega_n,\Omega_n^{\mathrm{ideal}}),
\qquad
T(\tau,\sigma)=\frac12\|\tau-\sigma\|_1.
\ee
This error is independent of the choice of purification.
Fix a purification whose reference has dimension
$\operatorname{rank}(\omega_{A^nB^n})$. Every other purification
is obtained from it by an isometry on the reference. This
isometry acts in the same way on the actual and ideal outputs
and preserves their trace distance.

\section{Universal state merging for weakly almost iid sources}

In this section we prove our main result, namely that the optimal
asymptotic net entanglement cost per pair for universal state
merging of weakly almost iid sources along a fixed reference
state $\rho_{AB}$ is $S(A|B)_\rho$.

\begin{thm}[(Universal state merging for weakly almost iid sources)]
\label{thm:universal-waiid-merging}
Fix a finite-dimensional bipartite state $\rho_{AB}$ and a rate
$R>S(A|B)_\rho$. There exists a sequence of one-way LOCC
state-merging protocols, depending only on $\rho_{AB}$ and $R$,
with net entanglement cost at most $nR$ ebits for all sufficiently
large $n$. The same sequence of protocols achieves vanishing
state merging error for every weakly almost iid source
$(\omega_n)_n$ along $\rho_{AB}$ and is thus universal for this class.

The optimal asymptotic net entanglement cost per copy for
universal merging of weakly almost iid sources along
$\rho_{AB}$ is $S(A|B)_\rho$.
\end{thm}

\begin{rem}
    Since every MSR almost iid source and every Wasserstein almost
iid source along $\rho_{AB}$ is weakly almost iid along
$\rho_{AB}$, the same sequence of protocols works for both
classes. Each class contains the exact iid source
$(\rho_{AB}^{\otimes n})_n$, so the optimal asymptotic net
entanglement cost per copy for universal merging is
$S(A|B)_\rho$ for each class.
\end{rem}

\begin{rem}
The proof shows that the same sequence of protocols achieves
vanishing state merging error for every sequence of states $(\omega_n)_n$
satisfying
\bb\label{eq:merging-two-pair-condition}
\lim_{n\to\infty}
\bigl\|(\bar\omega_n)_{12}-\rho_{AB}^{\otimes 2}\bigr\|_1=0,
\ee
where $\bar\omega_n$ is the average of $\omega_n$ over
permutations of the $n$ pairs $A_iB_i$, and
$(\bar\omega_n)_{12}$ is its marginal on the first two pairs.
\end{rem}

\begin{proof}
\vspace{0.1cm}

\medskip
\noindent\textbf{Protocol construction:}\quad
Write $\rho=\rho_{AB}$, $s=S(A|B)_\rho$, and
$d=\dim(\mathcal H_A\otimes\mathcal H_B)$. Fix $r$ with
$s<r<R$. By Lemma~\ref{lem:exponential-iid-merging},
there exist $\kappa>0$ and $n_0$ such that, for every
$n\ge n_0$, there is a single one-way LOCC protocol that
merges $\theta^{\otimes n}$ for every state $\theta$ with
$S(A|B)_\theta\le (s+r)/2$, at net entanglement cost at
most $nr$ and with state merging error at most
$\frac12 2^{-\kappa n}$. Fix one such protocol $\mathcal M_n$ for each $n\ge n_0$, with
initial and final Schmidt ranks $K_n$ and $L_n$, respectively.

{For the error analysis, with} $\kappa>0$ fixed,
choose $q\in(0,1)$ sufficiently small that the depolarised state
\bb
\rho_q=(1-q)\rho+qI/d
\ee
satisfies
\bb
S(A|B)_{\rho_q}\le \frac{s+r}{2}
\qquad\text{and}\qquad
-\log(1-q)<\frac{\kappa}{2}.
\ee
Such a choice is possible because
{$S(A|B)_{\rho_q}\to S(A|B)_\rho=s$} by continuity
and $-\log(1-q)\to 0$ as $q\downarrow 0$.
Since $\rho$ and $\rho_q$ are diagonal in a common eigenbasis
and each eigenvalue of $\rho_q$ is at least $1-q$ times the
corresponding eigenvalue of $\rho$, we have
\bb
D(\rho\Vert\rho_q)\le -\log(1-q)<\frac{\kappa}{2}.
\ee
The exponential merging bound therefore applies to
$\rho_q^{\otimes n}$ with the already fixed exponent
$\kappa$.\footnote{Applying the fixed-source result of
Horodecki, Oppenheim and Winter~\cite{HOW} separately to
$\rho_q^{\otimes n}$ for each sufficiently small $q$ gives a
positive exponent $\kappa(q)$ that may depend on $q$. This
would suffice if $D(\rho\Vert\rho_q)<\kappa(q)$ for some $q$,
since one could then choose $\eta>0$ with
$D(\rho\Vert\rho_q)+\eta<\kappa(q)$. However, knowing that
$\kappa(q)>0$ for each fixed $q$ and that
$D(\rho\Vert\rho_q)\to 0$ as $q\to 0$ does not guarantee
that there is a $q$ for which $D(\rho\Vert\rho_q)<\kappa(q)$.
The uniform bound in Lemma~\ref{lem:exponential-iid-merging}
allows us to fix $\kappa$ before choosing $q$.} We use $\rho_q^{\otimes n}$ as a \emph{comparison input} to
analyse the merging error on $\omega_n$. The domination lemma
bounds a high-weight compression of $\bar\omega_n$ by a multiple
of this full-rank tensor power.

Alice chooses a uniformly random permutation $\pi\in S_n$
and sends it to Bob\footnote{Averaging over $\pi$ makes the
state merging error bound depend on the permutation-averaged source
state $\bar\omega_n$, to which
Lemma~\ref{lem:product-state-domination} applies.}.
Alice and Bob each reorder their $n$ registers according to
$\pi$ and then run the fixed iid merging protocol with
the initial resource $\Phi_{K_n}^{A_0B_0}$.
{Alice applies the local instrument specified by
$\mathcal M_n$ to $A^nA_0$ and sends its outcome $x$ to Bob,
who applies the corresponding decoder to $B^nB_0$.} 
Bob then applies the inverse permutation to both $\widehat A^n$
and $B^n$, restoring the original order. Let $\Omega_{n,\pi}$
denote the output conditioned on Alice choosing $\pi$, after
Bob has restored the original order. When the parties discard
the classical record of $\pi$, the final output is the uniform
average
\bb
\Omega_n=\frac{1}{n!}\sum_{\pi\in S_n}\Omega_{n,\pi}.
\ee
The random permutation and its reversal require only local
unitaries and a classical message from Alice to Bob. The
overall protocol is therefore one-way LOCC and has the same
net entanglement cost as the fixed iid merging protocol,
namely $\log K_n-\log L_n\le nr\le nR$.

\medskip
\noindent\textbf{An error bound for arbitrary inputs:}\quad
For each $n\ge n_0$, let
\bb
V_n:\mathcal H_{A^nB^nA_0B_0}
\longrightarrow
\mathcal H_{\widehat A^n B^n A_1B_1E_n}
\ee
be a Stinespring isometry for the fixed iid merging
protocol. Tracing out $E_n$ gives the state on Bob's output
registers $\widehat A^n$ and $B^n$ and on the final
entanglement registers $A_1B_1$.

Let $W_n$ be the ideal merging isometry defined in
Section~\ref{subsec:state-merging} for the initial resource
$\Phi_{K_n}^{A_0B_0}$ and the prescribed final resource
$\Phi_{L_n}^{A_1B_1}$ of the fixed protocol.
If necessary, enlarge $E_n$ by appending a fixed pure ancilla
to the output of $V_n$. Embed the ideal environment $G_n$
isometrically into $E_n$ and continue to denote the resulting
isometry by $W_n$. The two isometries $V_n$ and $W_n$ now
have the same codomain. On a purified input, we extend both
isometries by the identity on the purifying reference $Z$.

Take a purification $\lvert\psi_q\rangle_{A^nB^nZ}$ of
$\rho_q^{\otimes n}$ and set
\bb
\lvert\chi_q\rangle
=
\lvert\psi_q\rangle_{A^nB^nZ}
\otimes\lvert\Phi_{K_n}\rangle_{A_0B_0}.
\ee
The vectors $(V_n\otimes I_Z)\lvert\chi_q\rangle$ and
$(W_n\otimes I_Z)\lvert\chi_q\rangle$ purify the actual
and ideal outputs, respectively, on $\widehat A^n B^n A_1B_1Z$.
The trace distance between these outputs is at most
$\frac12 2^{-\kappa n}$. By the Fuchs--van de Graaf inequality
$1-F(\tau,\sigma)\le T(\tau,\sigma)$~\cite{FuchsVanDeGraaf1999},
their fidelity $F_{n,q}$ is at least $1-\frac12 2^{-\kappa n}$.

By Uhlmann's theorem~\cite{Uhlmann-fidelity}, there is a
unitary $U_n$ on $E_n$ such that, with
$W_n'=(I\otimes U_n)W_n$, the overlap between
$(V_n\otimes I_Z)\lvert\chi_q\rangle$ and
$(W_n'\otimes I_Z)\lvert\chi_q\rangle$ is real and equal
to $F_{n,q}$. Since $U_n$ acts only on the environment,
tracing out $E_n$ gives the same ideal output for $W_n'$
as for $W_n$.

{
Fix this choice of $U_n$ for the error analysis. To use the
domination lemma, we express the squared distance between the
dilated output vectors as the expectation of a fixed positive
operator on the input space. For a state $\xi$ on $A^nB^n$, let
$\lvert\chi_\xi\rangle$ be a purification of $\xi$ on $A^nB^nZ$
tensored with $\lvert\Phi_{K_n}\rangle_{A_0B_0}$. Then
\bb
\begin{aligned}
\bigl\|[(V_n-W'_n)\otimes I_Z]\lvert\chi_\xi\rangle\bigr\|^2
&=\langle\chi_\xi\rvert
\Bigl[\underbrace{(V_n-W'_n)^\dagger(V_n-W'_n)}_{X_n}
       \otimes I_Z\Bigr]\lvert\chi_\xi\rangle\\
&=\operatorname{Tr}\!\left[
 X_n(\xi\otimes\Phi_{K_n}^{A_0B_0})\right].
\end{aligned}
\ee
The last equality uses the marginal of $\lvert\chi_\xi\rangle$
on $A^nB^nA_0B_0$. Since $V_n$ and $W'_n$ are isometries,
$\|V_n-W'_n\|_\infty\le 2$ and hence $0\le X_n\le 4I$.

For $\xi=\rho_q^{\otimes n}$, take
$\lvert\chi_\xi\rangle=\lvert\chi_q\rangle$. By the choice of $U_n$,
the two dilated output vectors have real overlap $F_{n,q}$, so
\bb\label{eq:merging-iid-error-operator}
\operatorname{Tr}\!\left[
 X_n(\rho_q^{\otimes n}\otimes\Phi_{K_n}^{A_0B_0})\right]
= 2-2F_{n,q} \le 2^{-\kappa n}.
\ee
For arbitrary $\xi$, the trace distance between the two
dilated pure states is at most the norm of the difference
between their vectors.
Contractivity under tracing out $E_n$ therefore bounds the
merging error of the fixed protocol on $\xi$ by
\bb
\sqrt{\operatorname{Tr}\!\left[
 X_n(\xi\otimes\Phi_{K_n}^{A_0B_0})\right]}.
\ee
}

\medskip
\noindent\textbf{Permutation averaging and domination:}\quad
Write $V_\pi$ for the unitary that permutes the $n$ pairs
in $A^nB^n$. Conditioned on Alice choosing $\pi$, the
source marginal presented to the fixed protocol is
$V_\pi\omega_n V_\pi^\dagger$. Bob's inverse permutation
restores the same ideal output $\Omega_n^{\mathrm{ideal}}$
for every choice of $\pi$ and preserves the trace distance
between the actual and ideal outputs. Thus convexity of
trace distance and the Cauchy--Schwarz inequality give
\bb
\begin{aligned}
T(\Omega_n,\Omega_n^{\mathrm{ideal}})
&\le \frac{1}{n!}\sum_{\pi\in S_n}
\sqrt{\operatorname{Tr}\!\left[
X_n\bigl((V_\pi\omega_nV_\pi^\dagger)
          \otimes\Phi_{K_n}^{A_0B_0}\bigr)
\right]}\\
&\le \sqrt{\operatorname{Tr}(X_n\sigma_n)},
\end{aligned}
\ee
where $\sigma_n=\bar\omega_n\otimes\Phi_{K_n}^{A_0B_0}$
and $\bar\omega_n$ is the permutation average of $\omega_n$.

{Suppose now that $(\omega_n)_n$ satisfies the
two-pair marginal condition~\eqref{eq:merging-two-pair-condition}.}
Choose $\eta>0$ such that $D(\rho\Vert\rho_q)+\eta<\kappa$.
Lemma~\ref{lem:product-state-domination} gives orthogonal projectors \(Q_n\)
on $A^nB^n$ such that
\bb
b_n:=1-\operatorname{Tr}(Q_n\bar\omega_n)\longrightarrow 0
\ee
and
\bb
Q_n\bar\omega_n Q_n\le C_n\rho_q^{\otimes n},
\qquad
C_n=(n+d)^{c_d}2^{n[D(\rho\Vert\rho_q)+\eta]},
\qquad
c_d=\frac{(d-1)(d+2)}{2}.
\ee
Let $\widetilde Q_n=Q_n\otimes I_{A_0B_0}$. Tensoring
the domination inequality with $\Phi_{K_n}^{A_0B_0}$
gives
\bb
\widetilde Q_n\sigma_n\widetilde Q_n
\le C_n\bigl(\rho_q^{\otimes n}\otimes\Phi_{K_n}^{A_0B_0}\bigr).
\ee
Since $X_n\ge 0$, equation~\eqref{eq:merging-iid-error-operator}
then yields
\bb
\operatorname{Tr}(X_n\widetilde Q_n\sigma_n\widetilde Q_n)
\le C_n 2^{-\kappa n}.
\ee

To bound $\operatorname{Tr}(X_n\sigma_n)$ using the estimate
on $\widetilde Q_n\sigma_n\widetilde Q_n$, take a purification
$\lvert\chi_n\rangle$ of $\sigma_n$.
In the following vector norms, operators on the input
are extended by the identity on the purifying reference.
Inserting $I=\widetilde Q_n+(I-\widetilde Q_n)$ and using
the triangle inequality gives
\bb
\begin{aligned}
\sqrt{\operatorname{Tr}(X_n\sigma_n)}
&=\bigl\|X_n^{1/2}\lvert\chi_n\rangle\bigr\|\\
&\le
\bigl\|X_n^{1/2}\widetilde Q_n\lvert\chi_n\rangle\bigr\|
+
\bigl\|X_n^{1/2}(I-\widetilde Q_n)
        \lvert\chi_n\rangle\bigr\|\\
&\le
\sqrt{\operatorname{Tr}(X_n\widetilde Q_n\sigma_n\widetilde Q_n)}
+2\sqrt{b_n}\\
&\le
(n+d)^{c_d/2}
2^{-\frac n2[\kappa-D(\rho\Vert\rho_q)-\eta]}
+2\sqrt{b_n}
\longrightarrow 0.
\end{aligned}
\ee
Indeed, $\|(I-\widetilde Q_n)\lvert\chi_n\rangle\|^2=b_n$,
and $X_n\le 4I$ bounds the norm after applying $X_n^{1/2}$.
The first term in the last line tends to zero because
$\kappa-D(\rho\Vert\rho_q)-\eta>0$, and the second tends
to zero because $b_n\to 0$. The projectors $Q_n$ enter
only in the error analysis.

{This proves vanishing merging error under the
two-pair marginal condition~\eqref{eq:merging-two-pair-condition}\footnote{The merging error measures the trace distance between
the actual and ideal joint outputs on $\widehat A^n B^n A_1B_1Z$.
By contractivity of trace distance under partial trace over
Bob's registers, vanishing merging error implies that the
state on $A_1Z$ approaches the ideal product marginal
$\operatorname{Tr}_{B_1}(\Phi_{L_n}^{A_1B_1})\otimes\psi_Z$.
Thus Alice's final register $A_1$ is asymptotically decoupled
from the reference $Z$.}.
As shown in the remark following
Lemma~\ref{lem:product-state-domination}, every weakly almost
iid source along $\rho_{AB}$ satisfies this condition.}
The protocol sequence was fixed using only $\rho_{AB}$ and $R$,
so it is universal for this class.

\medskip
\noindent\textbf{Optimality:}\quad
The class contains the exact iid sequence $(\rho^{\otimes n})_n$.
On this input, vanishing trace-distance error implies vanishing
purified-distance error by the Fuchs--van de Graaf inequalities.
Fix an error tolerance $\varepsilon>0$ small enough that the
smoothing parameter in the one-shot
converse~\cite[Theorem~5.3]{DupuisBertaWullschlegerRenner2014}
lies in $(0,1)$. For all sufficiently large $n$, the
purified-distance error is at most $\varepsilon$, so the
converse applies with an additive term independent of $n$.
Dividing the bound by $n$ and applying the iid asymptotic
equipartition property~\cite[Lemma~2.8]{DupuisBertaWullschlegerRenner2014}
gives the lower bound $S(A|B)_\rho$ on the asymptotic net
entanglement cost per pair. Together with achievability for
every $R>S(A|B)_\rho$, this establishes $S(A|B)_\rho$ as the
optimal asymptotic net entanglement cost per pair for universal
state merging of weakly almost iid sources along $\rho_{AB}$.
\end{proof}

{\subsection{A uniform error bound for iid merging}
\label{subsec:uniform-iid-merging}

The proof of Theorem~\ref{thm:universal-waiid-merging} uses
an error exponent $\kappa$ that is uniform over a family of iid
inputs. This allows us to fix $\kappa$ before choosing the
depolarisation parameter $q$.

\begin{lemma}
\label{lem:exponential-iid-merging}
Fix finite-dimensional systems $A$ and $B$ and real numbers
$s<r$. Suppose the family of states $\theta_{AB}$ satisfying
$S(A|B)_\theta\le (s+r)/2$ is nonempty. There exist $\kappa>0$
and $N_0$ such that, for each $N\ge N_0$, a single one-way
LOCC protocol has net entanglement cost at most $Nr$ ebits
and merges $\theta_{AB}^{\otimes N}$ with trace-distance error
at most $\frac12 2^{-\kappa N}$ for every $\theta_{AB}$ in
this family.
\end{lemma}

\begin{proof}
Let
\bb
\mathcal X
=
\left\{
\theta_{AB}\in\mathcal D(\mathcal H_A\otimes\mathcal H_B)
\;\middle|\;
S(A|B)_\theta\le \frac{s+r}{2}
\right\}.
\ee
Set $\delta=(r-s)/4$. By Proposition~10 of
Boche and Jan{\ss}en~\cite{BocheJanssen2014}, there are
$c>0$ and an integer $N_1$ such that, for each $N\ge N_1$,
a single one-way LOCC merging protocol works for all
$\theta\in\mathcal X$ with net entanglement cost at most
\bb
N\left(\sup_{\theta\in\mathcal X}S(A|B)_\theta+\delta\right)
\le N\left(r-\frac{r-s}{4}\right)<Nr.
\ee

For each $\theta\in\mathcal X$, let $\tau$ and $\sigma$ be the
actual and ideal merging outputs for a purification of
$\theta_{AB}^{\otimes N}$, with the purifying reference included
in both outputs. Boche and Jan{\ss}en use squared fidelity,
so their uniform guarantee reads $F(\tau,\sigma)^2\ge 1-2^{-cN}$
in our notation. The Fuchs--van de Graaf
inequality~\cite{FuchsVanDeGraaf1999} gives
\bb
T(\tau,\sigma)
\le \sqrt{1-F(\tau,\sigma)^2}
\le 2^{-cN/2}.
\ee
Set $\kappa=c/4$ and take
$N_0=\max\{N_1,\lceil 1/\kappa\rceil\}$.
Then $\kappa N\ge 1$ for every $N\ge N_0$, so
\bb
2^{-cN/2}=2^{-2\kappa N}\le \frac12 2^{-\kappa N}.
\ee
\end{proof}}

\subsection*{Acknowledgements}

The author thanks Andreas Winter for the question that prompted
this paper. After her talk at a workshop at the Isaac Newton
Institute for Mathematical Sciences, he asked whether the iid
entanglement cost of state merging remains achievable for weakly
almost iid sources. She is also grateful to Ryuji Takagi for pointing her to the
second version of the paper by Rizzo et al.~\cite{RizzoEtAl2026}.
The operator domination bound
(Lemma~\ref{lem:product-state-domination}), a key ingredient
in this paper, builds on her concentration argument in
Lemma~3 of~\cite{datta2026entropy}. Its proof uses two estimates
from Rizzo et al., namely their Schur-spectrum concentration
estimate and their lower bound on the dimensions of irreducible
permutation representations. 
The author thanks the Isaac Newton Institute for Mathematical
Sciences for its hospitality during the workshop. She acknowledges
support from the Engineering and Physical Sciences Research Council
(grant EP/Y028732/1).
\medskip

\noindent
\textbf{Statement on AI use.}
The author independently proved Lemma~\ref{lem:product-state-domination}
using two estimates from Rizzo et al.~\cite{RizzoEtAl2026}.
ChatGPT (GPT-6.2 Sol) pointed her to Proposition~10 of Boche
and Jan{\ss}en~\cite{BocheJanssen2014}. She derived the uniform
iid merging bound in Lemma~\ref{lem:exponential-iid-merging}
from this proposition and used it in the achievability proof
of Theorem~\ref{thm:universal-waiid-merging}.
She also used ChatGPT (GPT-6.2 Sol) to help prepare and revise the manuscript. The author finalised the organisation and presentation of the results and assumes full responsibility for the manuscript.

\bibliography{biblio}
\end{document}